\documentclass[graybox]{svmult}

\usepackage{mathptmx}       
\usepackage{helvet}         
\usepackage{courier}        
\usepackage{type1cm}        
\usepackage{makeidx}         
\usepackage{graphicx}        
\usepackage{multicol}        
\usepackage[bottom]{footmisc}
\usepackage[intlimits]{amsmath}
\usepackage{rotating}
\usepackage{booktabs}
\usepackage{multirow}
\usepackage{url}

\makeindex             

\begin{document}

\title*{On Modelling of Crude Oil Futures in a Bivariate State-Space Framework}
\author{Peilun He, Karol Binkowski, Nino Kordzakhia, Pavel Shevchenko}
\institute{Peilun He \at Macquarie University, NSW 2109, Australia, \email{peilun.he@students.mq.edu.au}
\and Karol Binkowski \at Macquarie University, NSW 2109, Australia,  \email{karol.binkowski@mq.edu.au}
\and Nino Kordzakhia \at Macquarie University, NSW 2109, Australia,  \email{nino.kordzakhia@mq.edu.au}
\and Pavel Shevchenko \at Macquarie University, NSW 2109, Australia, \email{pavel.shevchenko@mq.edu.au}}
%
%
\maketitle

\abstract{We study a bivariate latent factor model for the pricing of commodity futures. The two unobservable state variables representing the short and long term factors are modelled as Ornstein-Uhlenbeck (OU) processes. The Kalman Filter (KF) algorithm has been implemented to estimate the unobservable factors as well as  unknown model parameters. The estimates of model parameters were obtained by maximising a Gaussian likelihood function. The algorithm has been applied to WTI Crude Oil NYMEX futures data\footnote{Data provided by Datascope - \url{https://hosted.datascope.reuters.com}}.}

\section{Introduction}
\label{sec:introduction}

In this paper, the OU two-factor model is used for modelling of short and long equilibrium commodity spot price levels. Our motivation is driven by the development of a robust KF algorithm which will be used for joint estimation of the model parameters and the state variables. In a different setup, the parameter estimation problem for bivariate OU process using KF has been studied in \cite{favetto2010parameter} and \cite{KutoyantsYuryA.2019Opeo}.

In \cite{ChengBenjamin2018Polc} the KF is used to study the effect of stochastic volatility and interest rates on the commodity spot prices using the market prices of long-dated futures and options. In \cite{peters2013calibration} the Kalman technique has been applied to calibration, jointly with filtering, of partially unobservable processes using particle Markov Chain Monte Carlo approach. The extended KF was developed in \cite{ewald2019calibration} for estimation of the state variables in the two-factor model from \cite{schwartz1997stochastic} for the commodity spot price and its convenience yield.

In Sect.~\ref{sec:two_factor_model}, we will derive the linear partially observable system specific for commodity futures prices developed in the two-factor model, which represents an extension of \cite{schwartz2000short}, in the risk-neutral setting. In Sect.~\ref{sec:realdata}, the model will be applied to WTI Crude Oil NYMEX futures prices over 2001-2005, 2005-2009 and 2014-2018 time periods.

\section{Two-Factor Model with Risk Premium Parameters}\label{sec:two_factor_model}
We propose the two-factor model of pricing of commodity futures which represents an extension of \cite{schwartz2000short}, where the spot price $S_t$ is modelled as the sum of two unobservable factors $\chi_t$ and $\xi_t$,
\begin{equation}
\log(S_t) = \chi_t + \xi_t,
\label{eq:spot}
\end{equation}
where $\chi_t$ is the short-term fluctuation in prices and $\xi_t$ is the long-term equilibrium price level. We assume that  $\chi_{t}$ follows an OU equation and its expected value converges to 0 as $ t \to \infty  $,
\begin{equation}
d\chi_{t} = (-\kappa\chi_{t} - \lambda_{\chi})dt + \sigma_{\chi}dZ_{t}^{\chi},  \; \;  \kappa>0.
\label{eq:chi_rn}
\end{equation}
The changes in the equilibrium level of $\xi_t$ are expected to persist and $\xi_t$ is also assumed to be a stationary OU process
\begin{equation}
d\xi_{t} = (\mu_{\xi} - \gamma\xi_{t}-\lambda_{\xi})dt + \sigma_{\xi}dZ_{t}^{\xi},  \; \;  \gamma>0,
\label{eq:xi_rn}
\end{equation}
where $(Z_{t}^{\chi})_{t \ge 0}$ and $(Z_{t}^{\xi})_{t \ge 0}$ are correlated standard Brownian motions processes with $E(dZ_{t}^{\chi}dZ_{t}^{\xi}) = \rho_{\chi\xi}dt$;  $\sigma_{\chi}$ and $\sigma_{\xi}$ are the volatilities; $\gamma$ and $\kappa$ are the speed of mean-reversion parameters of $\chi$ and $\xi$ processes respectively; $(\mu_{\xi}-\lambda_{\xi})/\gamma $ is a long-run mean for $\xi$. In \cite{schwartz2000short}, only one factor had a mean-reverting property. In this work, both $\chi_t$ and $\xi_t$ are modelled as the mean-reverting processes. The parameters $\lambda_{\chi}$ and $\lambda_{\xi}$ in \eqref{eq:chi_rn} and \eqref{eq:xi_rn} were introduced as adjustments for market price of risk. The approach stems from the risk-neutral futures pricing theory developed in \cite{black1976pricing}. Given the initial values $\chi_0$ and $\xi_0$, $\chi_t$ and $\xi_t$ are jointly normally distributed. Therefore the logarithm of the spot price, which is the sum of $\chi$ and $\xi$, is normally distributed. Hence, the spot price is log-normally distributed and
\begin{equation}
\log[E^*(S_t)] = E^*[\log(S_t)] + \frac{1}{2}Var^*[\log(S_t)] = e^{-\kappa t}\chi_0 + e^{-\gamma t}\xi_0 + A(t),
\label{eq:logESRN}
\end{equation}
where $E^*(\cdot)$ and $Var^*(\cdot)$ represent the expectation and variance taken with respect to the risk-neutral distribution, and  
\begin{align}
A(t) =& -\frac{\lambda_{\chi}}{\kappa}(1 - e^{-\kappa t}) + \frac{\mu_{\xi} - \lambda_{\xi}}{\gamma}(1 - e^{-\gamma t}) \nonumber \\
&+ \frac{1}{2}\left(\frac{1 - e^{-2\kappa t}}{2\kappa}\sigma_{\chi}^2 + \frac{1 - e^{-2\gamma t}}{2\gamma}\sigma_{\xi}^2 + 2\frac{1 - e^{-(\kappa + \gamma)t}}{\kappa + \gamma}\sigma_{\chi}\sigma_{\xi}\rho_{\chi\xi}\right).
\label{eq:at}
\end{align}

Let $F_{0, T}$ be the current market price of the futures contract with maturity $T$.  For eliminating arbitrage,  the futures prices must be equal to the expected spot prices at the asset delivery time $T$. Hence, under the risk-neutral measure, we have \newline
$\log(F_{0, T}) = e^{-\kappa T}\chi_0 + e^{-\gamma T}\xi_0 + A(T).$
After discretization,  we will obtain the following AR(1) dynamics for bivariate state variable $x_t$
\begin{equation}
x_t = c + Gx_{t - 1} + w_t,
\label{eq:xt}
\end{equation}
where
$$x_t = \left[\begin{matrix} \chi_t \\ \xi_t \end{matrix} \right], c = \left[\begin{array}{c} 0 \\ \frac{\mu_{\xi}}{\gamma}(1 - e^{-\gamma \Delta t}) \end{array}\right], G = \left[\begin{array}{cc}e^{-\kappa \Delta t} & 0 \\ 0 & e^{-\gamma \Delta t} \end{array}\right], $$
and $w_t$ is a column vector of uncorrelated normally distributed
random variables with $E(w_t) = \textbf{0}$ and
$$Cov(w_t) = W = Cov[(\chi_{\Delta t}, \xi_{\Delta t})] = \left[\begin{array}{cc}
\frac{1 - e^{-2\kappa \Delta t}}{2\kappa}\sigma_{\chi}^2 & \frac{1 - e^{-(\kappa + \gamma) \Delta t}}{\kappa + \gamma}\sigma_{\chi}\sigma_{\xi}\rho_{\chi\xi} \\
\frac{1 - e^{-(\kappa + \gamma) \Delta t}}{\kappa + \gamma}\sigma_{\chi}\sigma_{\xi}\rho_{\chi\xi} & \frac{1 - e^{-2\gamma \Delta t}}{2\gamma}\sigma_{\xi}^2
\end{array}\right],$$
$\Delta t$ is the time step between $(t-1)$ and $t$. The relationship between the state variables and the observed futures prices is given by
\begin{equation}
y_t = d_t + F_t'x_t + v_t,
\label{eq:yt}
\end{equation}
where
$$y_t' = \left(\log(F_{t, T_1}), \dots, \log(F_{t, T_n})\right), d_t' = \left(A(T_1), \dots, A(T_n)\right),F_t = \left[ \begin{matrix} e^{-\kappa T_1}, \dots, e^{-\kappa T_n} \\ e^{-\gamma T_1}, \dots, e^{-\gamma T_n} \end{matrix} \right],$$
$v_t$ is  $n$-dimensional vector of uncorrelated normally distributed random variables, $E(v_t) = \textbf{0}$, $Cov(v_t) = V$ and  $T_1, \dots,T_n$ are the futures maturity times. In Sect.~\ref{sec:realdata}, we assume that $V$ is a diagonal matrix with non-zero diagonal entries $s=(s_1^2, s_2^2, \dots, s_2^2)$, i.e. the variance of the error term for the first contract is $s_1^2$ and $s_2^2$ for all other remaining contracts.  Let $\mathcal{F}_t$ be $\sigma$ - algebra generated by the futures contract up to time $t$. The prediction errors $e_t = y_t - E(y_t|\mathcal{F}_{t - 1})$ are supposed to be multivariate normally distributed, then the log-likelihood function of $y=(y_1, y_2, \dots, y_{n_T})$ can be written as
\begin{equation}
l(\theta; y) = -\frac{nn_T \log{2\pi}}{2} - \frac{1}{2}\sum_{t = 1}^{n_T}\left[\log{\left[\det(L_{t | t - 1})\right]} + e_t'L_{t | t - 1}^{-1}e_t\right],
\label{eq:loglikelihood}
\end{equation}
where the set of unknown parameters $\theta=(\kappa, \gamma, \mu_{\xi}, \sigma_{\chi}, \sigma_{\xi}, \rho_{\chi \xi}, \lambda_{\chi}, \lambda_{\xi}, s_1, s_2)$, $n_T$ is the number of time instances, $L_{t | t - 1} = Cov(e_t|\mathcal{F}_{t - 1})$. Given $y_t$, the maximum likelihood estimate (MLE) of $\theta$ is obtained by maximising the log-likelihood function from \eqref{eq:loglikelihood}. Both quantities  $e_t$ and $L_{t|t-1}$ are computed within the KF.

\section{Crude Oil Futures}\label{sec:realdata}

The unknown parameters were estimated \footnote{The Appendix containing the initial values and parameter estimates along with their standard errors can be found  at \url{https://github.com/peilun-he/MAF-Conference-September-2020}} by maximising the log-likelihood function \eqref{eq:loglikelihood}. Then, the state variables were estimated using the KF and Kalman Smoother (KS),  \cite{harvey1990forecasting} and \cite{de1989smoothing}. Given all observations until time $T$ and the current time $t$, $t \le T$, KF only uses the  observations up to $t$, while KS uses all the available observations up to $T$. In this section, ``in-sample'' and ``out-of-sample'' performances of KF and KS are analysed using the RMSE criterion.

We used the historical data of WTI Crude Oil NYMEX futures prices over different time intervals from 1996  to 2019. The data comprised the prices of 20 monthly futures contracts with duration up to 20 months.

\begin{figure}[ht]
	\centering
	\includegraphics[width=0.8\textwidth]{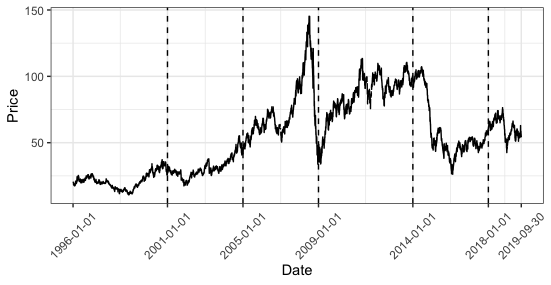}
	\caption{WTI Crude Oil futures prices of the first available contract.}
	\label{fig:clc1}
\end{figure}

Figure \ref{fig:clc1} shows the WTI Crude Oil futures prices from 1996 to 2019. It is obvious that the prices dropped dramatically during the Global Financial Crisis (GFC) in 2008. For studying the ``in-sample'' and ``out-of-sample'' forecasting performances, the three separate time periods were selected, 01/01/2001 - 01/01/2005, 01/01/2005 - 01/01/2009, and 01/01/2014 - 01/01/2018.

\begin{table}[ht]
	\caption{RMSE computed over three different time periods using two forecasting methods for each time interval.}\label{rmse_original}
	\centering
	\small
	\begin{tabular}{cccccccc}
		\hline\noalign{\smallskip}
		\multicolumn{2}{c}{Period} & \multicolumn{2}{c}{2001-2005} & \multicolumn{2}{c}{2005-2009} & \multicolumn{2}{c}{2014-2018} \\
		\noalign{\smallskip}\svhline\noalign{\smallskip}
		\multicolumn{2}{c}{Estimation} & Filter & Smoother & Filter & Smoother & Filter & Smoother \\
		\noalign{\smallskip}\svhline\noalign{\smallskip}
		\multirow{3}*{In-Sample} & C4 & 0.003264 & 0.003268 & 0.002219 & 0.002244 & 0.002180 & 0.002181 \\
		~ & C9 & 0.002289 & 0.002304 & 0.001612 & 0.001616 & 0.001646 & 0.001668 \\
		~ & C13 & 0.004155 & 0.004163 & 0.003959 & 0.003986 & 0.003516 & 0.003494 \\
		\noalign{\smallskip}\svhline\noalign{\smallskip}
		\multirow{2}*{Out-of-Sample} & C14 & 0.005959 & 0.005955 & 0.005569 & 0.005591 & 0.005215 & 0.005181 \\
		~ & C20 & 0.018585 & 0.018579 & 0.018002 & 0.018006 & 0.020331 & 0.020265 \\
		\noalign{\smallskip}\hline\noalign{\smallskip}
	\end{tabular}
\end{table}
 Table \ref{rmse_original} provides the RMSE over the selected time periods. ``In-sample'' forecasting performance has been evaluated on the first 13 contracts (C1-C13), while ``out-of-sample'' performance has been evaluated on the 14th to the 20th contracts (C14-C20). Overall, the ``in-sample" forecasting errors were less than ``out-of-sample" errors as seen in Table 1. The  ``out-of-sample" forecasting errors were consistently increasing with respect of maturity times from C14 to C20 contracts. The RMSE are consistent across the three time intervals, even over  2005 - 2009, where the futures prices plummeted during the GFC. In summary, for each specified time period, the RMSE calculated through KF is smaller for short maturity contracts, which provides evidence that the KF performed better in predicting prices for short maturity contracts, whilst KS outperformed KF in the pricing of longer maturity futures.

\begin{figure}[ht]
	\centering
	\includegraphics[width=0.9\textwidth]{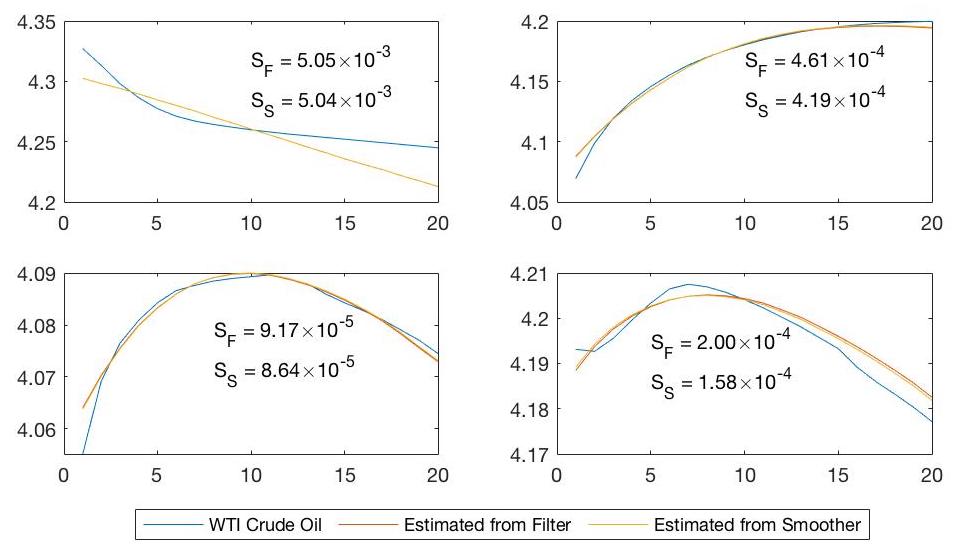}
	\caption{Cross-sectional graphs of the logarithms of futures prices and their forecasts on 4 different days. $S_F$ represents the sum of squares of estimation errors using KF; $S_S$ represents the sum of squares of estimation errors  using KS.}
	\label{fig:contango}
\end{figure}

Figure \ref{fig:contango} gives the cross-sectional data plots of the logarithms of futures prices and their forecasts obtained by KF and KS on four different days. The plots exhibit the distinct patterns of futures curves, 05/09/2007, 10/10/2006, 14/11/2005 and 20/09/2005. The horizontal axis represents the number of contracts from 1 to 20 and the logarithm of futures prices are presented on the vertical axis.  The RMSE for the curve with backwardation pattern (top left) appears larger than RMSE of that with contango pattern (top right)\footnote{Backwardation represents the situation where the futures prices with shorter maturities are higher than the futures prices with longer maturities, while contango refers to the reverse situation. }.



\section{Conclusion}\label{conclusion}

In this paper, we have developed the two-factor model which can be used for pricing of oil futures and forecasting their term structure which remains a most significant challenge, \cite{Cort2019}. The KF algorithm has been robustified by the grid-search add-on which has been implemented to estimate the hidden factors jointly with unknown model parameters. The model has been applied to WTI Crude Oil futures market prices from 1996 to 2019. The model ``in-sample'' and ``out-of-sample'' forecasting performances were evaluated using the RMSE criterion. Moreover, we observed that KF gives a better estimate of state vector $x_t$ for shorter maturity contracts, while KS performs better for contracts with longer maturities.

\end{document}